\documentstyle[prl,aps,floats]{revtex}

\begin{document}

\renewcommand{\topfraction}{0.8}
\twocolumn[\hsize\textwidth\columnwidth\hsize\csname
@twocolumnfalse\endcsname

\title{Spatial Variations of Fundamental Constants}
\author{John D. Barrow and Chris O'Toole \\
Astronomy Centre\\
University of Sussex\\
Brighton BN1 9QJ\\
UK}
\maketitle
\date{\today}

\begin{abstract}
We show that observational limits on the possible time variation of
constants of Nature are significantly affected by allowing for both space
and time variation. Bekenstein's generalisation of Maxwell's equations to
allow for cosmological variation of $\alpha $ is investigated in a universe
containing spherically symmetric inhomogeneities. The time variation of $%
\alpha $ is determined by the local matter density and hence limits obtained
in high-density geophysical environments are far more constraining than
those obtained at high redshift. This new feature is expected to be a
property of a wide class of theories for the variation of constants.

\vspace{0.3cm}
\end{abstract}

\vskip2pc]

There has been renewed interest in modelling and constraining possible time
variations in the traditional constants of Nature. New observational limits
have been stimulated by high-quality astronomical data \cite{webb}, whilst
theories of high-energy physics featuring additional dimensions of space and
new dilaton fields have provided motivations for studying variations in the
gravitational, strong, and electroweak coupling 'constants' \cite{string}, 
\cite{quin}. There exists a well-defined theoretical framework for studying
variations in the gravitational coupling by means of scalar-tensor gravity
theories and these have been studied in the case of purely time-dependent
variations arising in spatially homogeneous cosmological models and in the
static case. Different theoretical descriptions admitting variations in the
fine structure constant, $\alpha $, have been given by Bekenstein \cite{bek}
and by Barrow and Magueijo \cite{BM1}. Although it is usual to refer to all
these theories as descriptions of a variation in the value of a dimensional
constant (like $G,c,$ or $e$), only variations in dimensionless constants
have an invariant operational meaning under changes of units. Thus, it is
possible to find transformations which map the representation of a theory of
varying $\alpha $ into either a form with varying $c,$ \cite{AM}, or varying 
$e,$ \cite{BM2}.The published observational limits on the variations of
constants are derived from local geophysical and laboratory considerations 
\cite{oklo}, or from high-redshift astronomical observations of atomic
spectra \cite{webb}, \cite{astro}. Generally, the direct laboratory bounds 
\cite{pres} are significantly weaker than those derived from geophysics and
astrophysics because of the billions of years of look-back time over which
the latter two fields can gather data. These observations are traditionally
used to place limits on how much the value of a particular constant of
Nature can have varied over an interval of time. It is implicitly assumed
that the rate of time variation {\it is the same everywhere.}

In this paper we show why it is important to include the possibility of
spatial variations in the study of time-variations of physical constants.
Their inclusion is essential if local solar system or terrestrial limits on
the possible variation of constants are to be correctly correlated to
cosmological observations. Spatial variations in the values of constants
develop by analogy with spatial variations in the density, driven by the
process of gravitational instability as the universe expands. We show how
the two sets of variations are related in a particular theory which is most
expeditious for this purpose. This is the lagrangian theory for the
variation of the fine structure constant developed by Bekenstein \cite{bek}.

Bekenstein has formulated a theory with varying fine structure constant, $%
\alpha =e^2/\hbar c,$ defined by the requirements that at constant $\alpha $
it reduces locally to Maxwell's equations with minimal vector coupling; all
variations of $\alpha $ are dynamical and derivable from an action
principle; the action is causal, time symmetric, and gauge invariant; no
length scale smaller then the Planck length enters the theory; and the
gravitational field equations are unchanged. These requirements lead to an
additional contribution to the action

\[
S_\varepsilon =-\frac 12\hbar c\ell ^{-2}\int \varepsilon ^{-2}\varepsilon
_{,\mu }\varepsilon ^{,\mu }\sqrt{-g}d^4x 
\]
where the fundamental electron charge varies through a dimensionless
universal field $\varepsilon (x^\nu )$ defined by $e=e_0\varepsilon (x^\nu )$
with $e_0$ constant and $\ell $ is a constant length scale defining the
theory. The variation of this action leads to a propagation equation for $%
\varepsilon :$

\[
\Box \ln \varepsilon =\frac{\ell ^2}{\hbar c}\left[ \varepsilon \frac{%
\partial \sigma }{\partial \varepsilon }-\frac 1{8\pi }F^{\mu \nu }F_{\mu
\nu }\right] . 
\]
where $\Box$ is the covariant D'Alembertian. In the case of a spatially
homogeneous Friedmann background universe of any curvature, with expansion
scale factor $a(t),$ this simplifies to

\begin{equation}
(a^3\dot \varepsilon /\varepsilon )^{\cdot }=-\zeta (\frac{\ell ^2}{\hbar c}%
)\rho a^3c^4  \label{key}
\end{equation}
where dot denotes differentiation with respect to the comoving proper
cosmological time, $t$, $\rho $ is the matter density of the universe, and $%
\zeta \sim 10^{-2}$ is a dimensionless measure of the fraction of mass in
Coulomb energy for an average nucleon. Notice that in the dust-dominated era
of the universe the right-hand side of (\ref{key}) is constant since $\rho
\propto a^{-3}$. From now on we set $\hbar =c=1.$ The variation of the fine
structure is linked to that of $\varepsilon $ by $\dot \alpha /\alpha =2\dot
\varepsilon /\varepsilon .$

We can use eq. (\ref{key}) to investigate the effects of space and time
variations in $\varepsilon .$ Since the background cosmological evolution
for $a(t)$ is unchanged by introducing variations in $\varepsilon $ we can
exploit the Birkhoff property of the gravitational sector of the theory to
create an exact description of a universe possessing spherically symmetric
inhomogeneities in density, curvature and, electron charge by matching
together solutions of different spatial curvature and density. We can then
calculate the evolution of $\varepsilon $ in both regions and correlate them
with inhomogeneities in the density. The Friedmann equation describing the
evolution of a dust inhomogeneity with density parameter $\Omega _0$ and
Hubble parameter $H_0$ is

\begin{equation}
\dot a^2=\frac{\Omega _0H_0^2}a(1-Ba);B\equiv \frac{\Omega _0-1}{\Omega _0}
\label{fr1}
\end{equation}

Using (\ref{fr1}), we can integrate (\ref{key}), to obtain

\begin{equation}
\frac{\dot \varepsilon }\varepsilon =\frac C{B^{3/2}a^3}\sin ^{-1}(\sqrt{Ba}%
)-\frac C{Ba^{5/2}}(1-Ba)^{1/2}+\frac{a_c}{a^3},  \label{key2}
\end{equation}
where we have defined the constant

\[
C\equiv -\zeta \ \ell ^2\rho a^3\Omega _0^{-1/2}H_0^{-1}=-\frac{3\zeta }{%
8\pi }(\frac \ell {\ell _p})^2\Omega _0^{1/2}H_0, 
\]
where $\ell _p$ is the Planck length. We note that in the flat background
universe ($\Omega =1$) we have $B=0$, so (see also \cite{liv})

\[
\frac{\dot \varepsilon }\varepsilon =C\left[ \frac 2{3a^{3/2}}+\frac{a_c}{a^3%
}\right] \ \ 
\]
and

\begin{equation}
\ln \varepsilon =\frac{2C_{*}}3\left[ \ln a+a_c(1-a^{-3/2})\right]
\label{flat}
\end{equation}
where $a(t)=(t/t_0)^{2/3}$ and a boundary condition $\varepsilon (t_0)=1$ is
imposed, following \cite{bek}. Hence, we have

\[
\varepsilon =\left( \frac t{t_0}\right) ^{C_{*}}\exp [\frac{2C_0a_c}3\{1-(%
\frac{t_0}t)\}]. 
\]

In a closed universe, when $\Omega >1$ we integrate (\ref{key2}) to obtain a
generalisation of (\ref{flat}),

\begin{eqnarray}
\ln \varepsilon &=&\frac{2C_{*}}3\{\ln a+\frac
1{Ba}-a^{-3}(2Ba-1)(1-Ba)^{1/2}\times  \nonumber \\
&&\times \left[ c_1+B^{-3/2}\sin ^{-1}(\sqrt{Ba})\right] \},  \label{clos}
\end{eqnarray}
where $C_{*}=-\frac{3\zeta }{8\pi }(\frac \ell {\ell _p})^2$ and $c_1$ is a
constant.

These results form the basis of a quantitative model for the variation of $%
\varepsilon $ in time and space. We model an inhomogeneous universe
containing spherically symmetric curvature inhomogeneities by matching a
flat ($k=0$) Friedmann universe with scale factor $a(t)\equiv R(t)$
containing matter with zero pressure and density $\rho $ to a curved ($k\neq
0$) Friedmann universe with scale factor $a(t)\equiv S(t)$ and density $\rho
+\delta $ The standard metric matching conditions, $R=S$ and $dR/dt=dS/dt$,
are applied initially at $t=t_1$. The hydrostatic support is maintained in
the absence of pressure and allows the same time coordinate to be used in
both regions. From the conservation equation we have $\rho _sS^3=(1+\delta
)\rho R^3$ where $\rho _s$ is the density in the lump and $\rho $ is the
density in the flat background universe and we define $\delta =(\rho _s-\rho
)/\rho $ and $\delta _1\equiv \delta (t_1)$. The matching of the flat and
curved regions gives, for the overdense region

\[
\dot S^2=C\left[ (1+\delta _1)S^{-1}-\delta _1S_1^{-1}\right] 
\]
Integrating and using the Friedmann equation for the background evolution of 
$R(t)\propto t^{2/3}$, this gives the relation

\begin{eqnarray*}
\frac 23R^{3/2}(1+\delta _1)^{1/2} &=&C\ -S_{+}(S-\frac{S^2}{S_{+}})^{1/2}
 \\
&&\ \ \ -S_{+}^{3/2}\tan ^{-1}[(\frac{S_{+}}S-1)^{1/2}],
\end{eqnarray*}
where $S_{+}=S_1(1+\delta _1^{-1})$ is the expansion maximum in the
positively curved lump when $\Omega _0>1,$ and $C$ is a constant that can be
expressed in terms of $S_1$ and $\delta _1$ alone. When the density contrast
is small ($\delta <1$) we have $\delta \propto t^{2/3}$. Using (\ref{key}),
we have the following relationship linking the rate of change of the
electron charge in the background universe, $\varepsilon (t)$, to its
variation in a density inhomogeneity with contrast$,\delta ,$ described by $%
\varepsilon _s(t),$

\begin{equation}
\frac{\dot \varepsilon _s}{\varepsilon _s}=(1+\delta )\frac{\dot \varepsilon
\ }\varepsilon +\frac C{S^3}  \label{var}
\end{equation}
Eqn. (\ref{var}) can also be applied to the case of underdensities ('voids')
modeled by open Friedmann regions with $\Omega <1$ when the contrast
parameter lies in the range $-1<\delta <0$. It shows how variations in local
cosmological density influence are coupled to the time variation of $%
\varepsilon $. The effects are likely to be large when comparing intra and
extragalactic effects because a typical virialised galaxy has an overdensity
of $\delta \sim 10^6.$ The spatial distribution of density inhomogeneities
can be described by means of the 2-point correlation function $\xi (r)\simeq
(1Mpc/r)^{1.8}$ \cite{corr}$.$ Statistically homogeneous and isotropic
spatial variations in the statistics of $\dot \varepsilon /\varepsilon $
should therefore follow the same trend with

\[
<\frac{\dot \varepsilon _s}{\varepsilon _s}({\bf x}+{\bf r})\frac{\dot
\varepsilon _s}{\varepsilon _s}({\bf x})>\propto r^{-1.8} 
\]

The propagation equation (\ref{key}) has the form $\Box (\ln \varepsilon
)\propto \rho .$ The action of the D'Alembertian on $\ln \varepsilon $
ensures that the theory is hyperbolic with nice properties \cite{bek,bmag}.
Typically, we expect other theories incorporating varying 'constants'
derived from the variation of a scalar field to have a similar structure 
\cite{bmag}. In the case of Brans-Dicke theory we find a similar form, with,
the Brans-Dicke scalar, $\phi \sim G^{-1}$ governed by \cite{dicke}

\begin{equation}
\Box \phi =\frac{8\pi }{3+2\omega }T  \label{bd}
\end{equation}
where $T$ is the trace of the energy momentum tensor of the matter sources.
Again we see how there is direct coupling between the {\em local} density of
matter (on the right-hand side of eq. (\ref{bd})) and the variations of the
scalar field associated with $G$. The full system of Brans-Dicke field
equations is more complicated to solve than those of the Bekenstein theory
examined above but the same general structure is obtained. A bound
overdensity with large density contrast over the flat background universe
will possess a different value of $\phi $ and of $\dot \phi $ at any given
time compared to the values of these quantities in the background. In the
weak-field limit the Brans-Dicke cosmological equations have the decoupled
linear form for the evolution of the gravitational potential perturbation,$%
\Phi ,$the density excess over the background universe $\delta ,$ and $\Psi
=\delta \phi /\phi $ which gives the relative perturbation of the
Brans-Dicke parameter over the background $\phi (t)$, so $\delta =0$ when $%
\Psi =0,$

\[
a^{-2}\nabla ^2\Phi =-4\pi \left( \frac{4+2\omega }{3+2\omega }\right) \phi
^{-1}\delta 
\]

\begin{equation}
a^{-2}\nabla ^2\Psi =-8\pi \left( \frac{\ 1}{3+2\omega }\right) \phi
^{-1}\delta  \label{wf2}
\end{equation}

As in the Bekenstein theory, the magnitude of the variations in $\dot G/G$
from site to site in the universe will be governed by the variations in the
density. Eqn. (\ref{wf2}) shows how the density perturbations create
inhomogeneity in $\phi $ and hence in $G$ with $\delta G/G\sim \ \delta \phi
/\phi \sim \Phi /(2+\omega ).$

Existing observational evidence has assumed that there is a single global
rate of change in time of a constant, like $G$ or $\alpha $, which is the
same everywhere. Theories like Bekenstein's for varying $\alpha $ and
Brans-Dicke gravity, have direct coupling between the variation of the
scalar whose changes drive variations in the 'constant' and the local matter
density. The rate of change of the fine structure constant in Bekenstein's
theory will be proportional to the density of the environment where the
variation is being observed or constrained. From (\ref{var}) we see that the
rate of variation is closely linked to the local matter density. Thus a
constraint on time variation of the fine structure constant obtained by
examination of the requirement for a neutron-capture resonance to have been
in place $1.8\times 10^9$ years ago in the Oklo natural reactor under the
Earth's surface in Gabon, \cite{oklo}, where $\rho >1$ gmcm$^{-3}$
corresponds to a cosmological background limit that could be $10^{29}$ times
stronger. Likewise, when one infers upper limits on the possibility of
time-varying $G$ from observations in the solar system of the binary-pulsar
system they should not be compared directly with cosmological limits without
renormalisation for the effects of density enhancement due to the formation
of galaxies. Thus they should correspond to far stronger cosmological limits
when they have been scaled by the appropriate density ratio $\rho
(solarsystem)/\rho _0$, where $\rho _0$ is the background density of the
universe. These considerations should provoke a reassessment of the
relationships between different astrophysical and geological limits on
possible variations of constants. The overall result will be to dramatically
strengthen restrictions on allowed cosmological variations of the
traditional fundamental constants. Typically, terrestrial and Galactic
limits of time variation could be strengthened by a factor of order $\rho
(local)/\rho _0\sim 10^{29}$ or $\rho (galactic)/\rho _0\sim 10^6$ when
translated onto cosmological scales, depending on the details of the Galaxy
and planetary formation process. This has significant consequences for the
investigation of additional spatial dimensions that appear to be required in
some versions of string theory and M theory \cite{string} and some versions
of quintessence \cite{quin}. We hope that this will stimulate a detailed
analysis of the complex sequence of events that transform the
almost-homogeneous evolution of varying constants on an extragalactic
cosmological scale into the complicated but greatly amplified pattern of
variations that would be produced by gravitational collapse to the
high-density islands of matter in which and on which we reside and from
which we make our observations.

Acknowledgements. The authors are supported by the PPARC. We would like to
thank J. Magueijo for helpful discussions.

\end{document}